# Effects of different metal cations on conformational transitions in PolyrA·PolyrU under concentration and temperature changes

Blagoi Yu.P., Egupov S.A., Usenko E.L., Sorokin V.A.

B.I. Verkin Institute for Low Temperature Physics and Engineering, National Academy of Sciences of Ukraine, 47, Lenin Ave., Kharkov, 61103, Ukraine
Electronic address: blagoi@ilt.kharkov.ua

## Abstract

By the theory based on the equilibrium binding model, phase diagrams have been calculated for the poly rA·poly rU system in the presence of $Mg^{2+}$, $Ni^{2+}$, $Cd^{2+}$, with $Na^{+}$ 0.1 M. Binding constants ($K_A$, $K_U$, $K_{AU}$, $K_{A2U}$) have been determined, the best agreement with the early experimental results being observed. Calculations of diagrams with $Mg^{2+}$ are shown to permit obtaining satisfactory results if steady mean constants of ion binding to polynucleotides of all the mentioned structures are used. But in the cases with $Ni^{2+}$ and $Cd^{2+}$, constants dependences on ion concentrations must be taken into account, especially with high ion contents in the presence of which compaction of molecules of single-stranded poly A emerges. It has been revealed that the main factor being responsible for differences in diagrams for $Ni^{2+}$ and $Cd^{2+}$ is a significant variation of their constants of binding to poly A. Temperature dependences of conformational transitions obtained by UV spectroscopy for low polymer concentrations are compared with those obtained by IR and VCD spectroscopy at high contents of these molecules and, accordingly, metal ions in solution. Differences and peculiarities conditioning irreversibility of the earlier observed disproportional 2→3 transition are clarified.

## Introduction

By the present time, the great number of studies on conformational transitions in double and triple chains formed by homopolynucleotides poly A and poly U with different metal ions ($Mg^{2+}$, $Ni^{2+}$, $Cd^{2+}$ and others) has been carried out, some experimental methods (UV-, IR- and VCD-spectroscopy) being used [1-10]. Dependences of AU↔A+U, 2AU↔A2U+A and A2U↔A+2U on ions concentrations and temperature have been ascertained. Therefore, it is of interest to consider possibilities of applying existing model theories of helix-coil transitions for describing the results obtained. This permits the prediction of the transition character at different ion concentrations and the determination of constants of the ion binding to polymers.

In our previous works [1,2], the equilibrium binding model was successfully used, the model having been developed by Frank-Kamenetsky M.D. [11] for describing 2↔1 transitions ("melting") of heterogeneous native DNA in the presence of biologically active substances. With relation to the binding of monomeric substances, in this case both DNA strands are equivalent after their separation. Nevertheless, upon polymers binding to ions, mutual repulsion of the bound cations has to be taken into account under the increase of their concentration in solution and possible compaction of single-stranded molecules in the presence of the ions mentioned. The problem is more complicated in the case of synthetic homopolynucleotides forming three- and four-stranded helices consisting of purine and pyrimidine strands. In this case several transition types (2↔1, 3↔2, 2↔3, 3↔1, 4↔1) are known to be possible. As well, constants of ion binding to purine and pyrimidine strands and dependences of these constants on the binding degree and temperature differ significantly. Besides, the transition of polynucleotides into the compact form is observed, followed with significant changes of binding constants and with precipitation of compact particles. In particular, this process can provide an explanation for irreversibility of 2↔3 transition (2AU↔A2U+A) in solution with $Ni^{2+}$, mentioned in [9]. For calculations, all the stated requires the additional experimental and theoretical data on ion-polynucleotide complexes and can permit obtaining ambiguous calculation results.

In previous papers on $Mg^{2+}$, $Ni^{2+}$ and $Cd^{2+}$ complexes with AU and A2U the dependence of binding constants on ion concentrations has not been taken into account that is averaged values of constants K have been used, and the description was of the qualitative character. In the present work an attempt is made for more rigorous describing the experimental data for transitions in polynucleotide structures formed with poly rA and poly rU with $Mg^{2+}$, $Ni^{2+}$ and $Cd^{2+}$, all the above factors being taken into account.

## Approximation of melting phase diagrams for metal cations by the probable distribution method and determination of binding constants

Experimental dependences of melting temperatures of polynucleotides in solutions with metal cations can be described theoretically by formulas [11-14] based on the model of the most probable distribution. At the same time constants of the ion binding to polynucleotides of the appropriate structure can be determined. More generally, formulas for transitions are the following:

1) for 2↔1 transition

$$T_{m2\to1} = T_{02\to1} / (1 - \frac{1}{2} \cdot (R \cdot T_{02\to1} / \Delta H_{2\to1}) \cdot \ln((1 + K_{AU} \cdot A_f)^2 / ((1 + K_A \cdot A_f) \cdot (1 + K_U \cdot A_f)))) \quad (1)$$

2) for 3↔1 transition

$$T_{m3\to1} = T_{03\to1} / (1 - \frac{1}{3} \cdot (R \cdot T_{03\to1} / \Delta H_{3\to1}) \cdot \ln((1 + K_{A2U} \cdot A_f)^3 / ((1 + K_A \cdot A_f) \cdot (1 + K_U \cdot A_f)^2))) \quad (2)$$

3) for 2↔3 transition

$$T_{m2\to3} = T_{02\to3} / (1 - \frac{1}{2} \cdot (R \cdot T_{02\to3} / \Delta H_{2\to3}) \cdot \ln((1 + K_{AU} \cdot A_f)^2 / ((1 + K_{A2U} \cdot A_f) \cdot (1 + K_A \cdot A_f)))) \quad (3)$$

where $T_m$ is the melting temperature for the appropriate (proper) transition, $T_0$ is the asymptotic melting temperature at the zero concentration of cations, $\Delta H$ is enthalpy of the corresponding transition, R is the universal gas constant, $A_f$ being concentrations of free cations in solution, and $K_A$, $K_U$, $K_{AU}$, $K_{A2U}$ are constants of cation binding to certain polynucleotide structures.

To determine binding constants and their concentration dependences, we used the following two approaches:

1. In the simplest case (when concentration dependences of constants of the ion binding to polynucleotides were absent) the least-squares method was used for calculations. In this method constant values varied up to the moment when the mean square deviation of theoretical $T_m$ values from the experimental ones

$$S(K_A, K_U, K_{AU}, K_{A2U}) = \sum_k (T_{m_{theor}}(\log A_{f_k}) - T_{m_{exp}}(\log A_{f_k}))^2 \to \min$$

took on the minimum value.

Unfortunately, the method gave satisfactory results only for $Mg^{2+}$ ions for which concentration dependences of binding constants could not be taken into account. Reasons of the above will be discussed below.

2. We used experimental values and dependences of binding constants K (C) [7,8] on the binding degree, recounted them to the concentration $A_f$ by the formula

$$A_f = \frac{C}{(1-C) \cdot K(C)},$$

substituted $K(A_f)$ into formulas (1-3) and corrected their $K(A_f)$ values, comparing the calculated $T_m$ with their experimental values.

Within the framework of the first approach, values of the constants were obtained for three branches of diagrams for 2→1, 3→1, 2→3 systems of poly rA-rU with $Mg^{2+}$ (Table 2), parameter values being presented in Table 1.

Table 1. Parameters used in temperature calculations of transitions in systems of rA-rU with cations.

| Transition | ΔH, kal/mol | $B_{ij}$ | $T_o$, K |
|---|---|---|---|
| 2→1 | 9000 | 12,1 | |
| 3→1 | 12000 | 9,6 | 332±1 |
| 2→3 | 4000 | 25 | |

Table 2. Values of cation-polyribonucleotide binding constants, obtained upon least-squares method of experimental phase diagrams (their concentration dependences were not taken into account).

| катион | $Mg^{2+}$ | $Ni^{2+}$ | $Cd^{2+}$ |
|---|---|---|---|
| $K_A$ | 264 | 359 | 5583 |
| $K_U$ | 170 | 250 | 250 |
| $K_{AU}$ | 282 | 855 | 2342 |
| $K_{A2U}$ | 774 | $-123 \cdot \log A_f + 854$ (*) | |

(*) The dependence was obtained for rA-rU with $Mg^{2+}$ by the mean-square method at fixed $K_A$=200 and $K_U$=250 and varied $K_{AU}$=428.

Transition temperatures $T_m$ determined experimentally and theoretically (by formulas (1-3)) for this system are given in Fig. 1.

To estimate constant $K_A$ and $K_{AU}$ (by the order of their values) for rA-rU systems with $Ni^{2+}$ and $Cd^{2+}$, we varied them by the least-squares method. In this case basic parameters were applied (Table 1). At that the constant $K_U$ was accepted as fixed and equal to 250, and $K_{A2U}$ was considered as dependent on log $A_f$ with fixed coefficients: $K_{A2U} = -123 \cdot \log A_f + 854$, coefficients being calculated by the least-squares method for rA-rU system with $Mg^{2+}$ with the fixed $K_A$=200 and $K_U$=250 and varied $K_{AU}$=427.63. The constants obtained are given in Table 2. As is seen, $K_A$ and $K_{AU}$ for $Cd^{2+}$ are by the order higher than those for $Ni^{2+}$.

For more precise calculations in the case of $Ni^{2+}$ and $Cd^{2+}$ ions, the dependence of ion binding constants must be taken into account.

As is seen from formulas (1)-(3) for $T_m$ calculations, the data are necessary on transition enthalpies $\Delta H_{ij}$ and on ion binding constants being functions of temperature concentrations of the ions bound. The above formulas applied for describing experimental phase diagrams permit to determine these values for polynucleotides of different structures and to appreciate differences in the ions action. But it is seen that, with the great number of unknown magnitudes in formulas (1)-(3), in the general case the determination of precise ("correct") values can be ambiguous. Therefore, literature experimental data from some papers were used in our calculations by the above formulas, corrected and made more exact. Enthalpy values of transitions $\Delta H_{ij}$ were taken from experimental works of Stevens and Felsenfeld [15], Krakauer [16], Klump [17] and others. But if to take into account the known theoretical supposition that enthalpies of transitions in homopolynucleotides are conditioned with the degree of the chain disordering and the stacking failure in this case (Volkenshtein, Molecular biophysics [18]), these values are not dependent on the ion kinds.

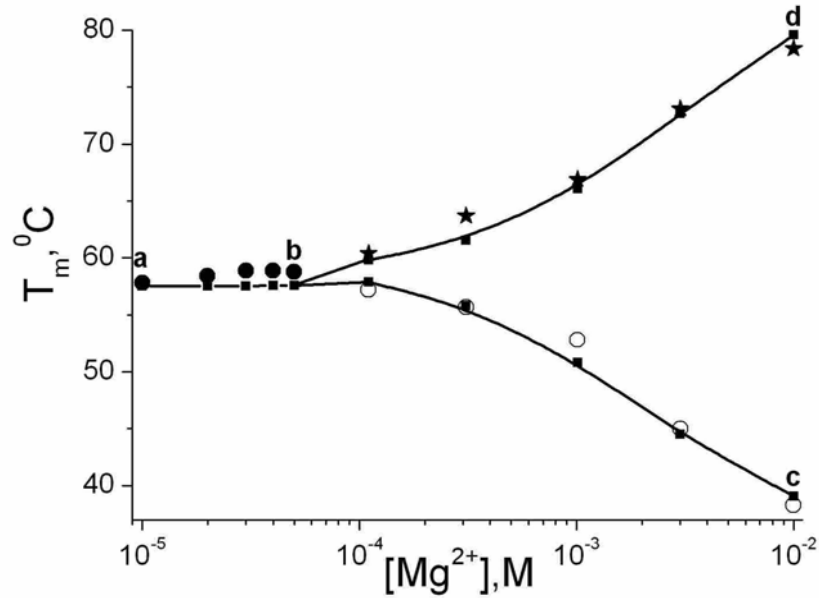

Figure 1. Phase diagram of rA·rU with $Mg^{2+}$.
Experimental $T_m$ values are shown by signs: ● for $(T_m)_{21}$, ○ - $(T_m)_{23}$, ★ for $(T_m)_{31}$. -■- calculation values.

$\Delta H$ values used (Table 1) are within dispersion of the data obtained by different authors at $T_m \sim (55-60)\ ^0C$. Besides, for simplification of calculations and for greater visualization of the results, formulas (1)-(3) were rearranged to

for 2→1 transition

$$\delta T_{21} = B_{21} \cdot \ln \frac{(1+K_{AU} \cdot A_f)^2}{(1+K_A \cdot A_f) \cdot (1+K_U \cdot A_f)} \qquad (4)$$

$$B_{21} = \frac{R \cdot T_0 \cdot T_{m21}}{2\Delta H_{21}}; \quad \Delta H_{21}=9000\ (kal/mol);$$

for 3→1 transition

$$\delta T_{31} = \frac{2}{3} \cdot B_{31} \cdot \ln \frac{(1+K_{A2U} \cdot A_f)^3}{(1+K_A \cdot A_f) \cdot (1+K_U \cdot A_f)^2}; \quad B_{31} = \frac{R \cdot T_0 \cdot T_{m31}}{2\Delta H_{31}}; \qquad (5)$$

$\Delta H_{31}=12000\ (kal/mol);$
for 2→3 transition

$$\delta T_{23} = B_{23} \cdot \ln \frac{(1 + K_{AU} \cdot A_f)^2}{(1 + K_{A2U} \cdot A_f) \cdot (1 + K_A \cdot A_f)}; \quad B_{23} = \frac{R \cdot T_0 \cdot T_{m23}}{2\Delta H_{23}} \quad (6)$$

$\Delta H_{23}$=4000 (kal/mol),
where $\delta T_{ij} = T_m - T_0$.

As $T_m \approx T_0$, $B_{ij} = \dfrac{R \cdot T_0^2}{2\Delta H_{ij}}$ was set to be used. For 2→3 and 3→1 transitions $T_0$ fits with the beginning point of these transitions.

## Calculation results and discussion

Calculation results of $\delta T_m$ dependences on ion concentrations are examined and compared with the experimental data (Figs. 1-3). This permits to ascertain the principal factors determining forms of phase diagrams for various ion types.

First of all, it should be noted that in the case of $Mg^{2+}$ ions it is possible to calculate diagrams by fixed constants independent on ion concentrations (Formulas 1-3). This may be conditioned with the fact that $Mg^{2+}$ ions (as well as $Na^+$ and $K^+$) bind only to negative charges of phosphate groups, not resulting in compaction of polymeric molecules. In this case binding constants for all the polynucleotide strands (ordinary, double and triple ones) decrease with the rise of ion concentrations, and ratios of these constants in formulas (4)-(6) keep about constant. Therefore, mean values of the constants used in calculations (Table 2) describe rather well the phase diagram (Fig. 1). $T_{32}$ value lowers and $T_{31}$ increases with the ion concentration rise. This calculation diagram agrees well with the experimental data and is similar to those for $Na^+$ and $K^+$ which interact too only with charges of phosphate groups. Values of binding constants for $Mg^{2+}$ ions (Table 2) differ slightly from earlier ones [4] and describe more exact the experimental data.

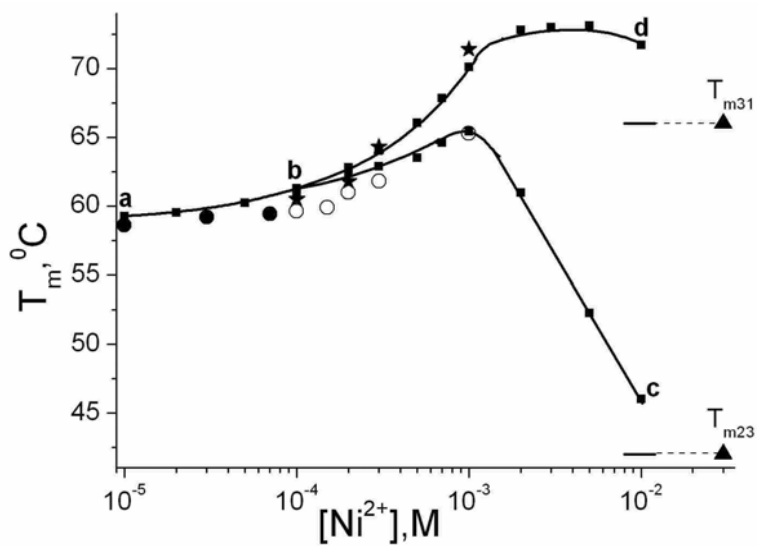

Figure 2. Phase diagram of rA·rU with $Ni^{2+}$.
Experimental $T_m$ values are shown by signs: ● for $(T_m)_{21}$, ○ - $(T_m)_{23}$, ★ for $(T_m)_{31}$. -■- calculation values.
▲ – $T_{m31}$ and $T_{m23}$ are transition temperatures obtained by IR and VCD methods at corresponding total ion concentrations in solution.
---- – $T_m$ values corresponding to approximated concentration $A_f$ of free ions, used in calculations.

More complicated concentration dependences of the transition temperatures $T_{31}$ and $T_{23}$ are observed for $Ni^{2+}$ and $Cd^{2+}$ (in contrast to $Mg^{2+}$). First, points of the beginning of disproportion transitions 2→3 shift to the region of higher ion concentrations and do not coincide with points of 3→1 transitions beginning for A2U system. Besides, at $Ni^{2+}$ concentration in the range to $10^{-3}$ M values of both $T_{31}$ and $T_{23}$ increase that is both the diagram branches are up-directed, and at higher $Ni^{2+}$ concentrations dispersion of UV irradiation is observed, evidencing the initiation of polymers compaction (presumably, of poly A), and this prevents from measurements by UV spectroscopy [3]. IR and VCD spectroscopy applied [9] permitted to determine $T_{23}$ and $T_{31}$ values at $A_f$ corresponding to $\sim 10^{-2}$M concentration of $Ni^{2+}$. At this concentration $K_A$ values reach $\sim 2 \cdot 10^3$ $M^{-1}$ magnitude because of the ion binding to bases in compact particles. In this case $\delta T_{23} \approx -11$ $^oC$ and $\delta T_{31} \approx +14$ $^oC$ are close to values for $Mg^{2+}$ (Fig. 2). It should be noted too that a higher concentration of poly A·poly U ($\sim 10^{-2}$M) shifts the beginning of disproportion transitions into the region of higher concentrations. Unfortunately, in this case correct calculations of $A_f$ for $Ni^{2+}$ are impossible because of lack of the data on constants of the ion binding to polynucleotides, especially if to take into account possible compaction of poly A. Therefore, Figure 2 presents points ▲ of 2→1 ($T_{m21}$) and 3→1 ($T_{m31}$) transitions at the total ion concentration $3 \cdot 10^{-2}$ M and A/polymer content P ratio being equal to 0.4. Regions of approximated $A_f$ values of corresponding transitions are marked with horizontal lines.

Unlike $Ni^{2+}$ ions, $Cd^{2+}$ reduces temperatures $T_{23}$ and $T_{31}$ (Fig. 3), $T_{23}$ decreasing especially sharply [10]. As seen from Figure, UV spectroscopic data on existence regions of double-, triple- and single-stranded structures correlate qualitatively with IR and VCD spectroscopic findings. As in the case of $Ni^{2+}$, it was impossible to perform exact quantitative comparison of $A_f$ values obtained by UV and IR spectroscopic methods because of high polynucleotide concentrations and molecule compaction. In this case the concentration of free ions ($A_f$) in theoretical calculations is approximate. The transition of poly A molecules into the compact state results in irreversibility of the disproportional transition 2→3 [3, 10]. At any case, during the experiment with cooling up to $T_m \sim 20\ ^0C$ the equilibrium in the system does not change practically.

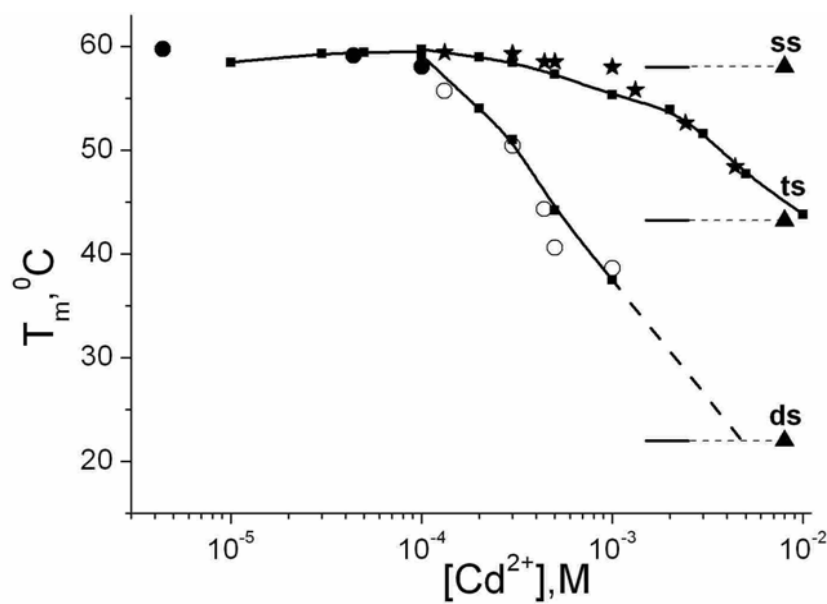

Figure 3. Phase diagram of rA·rU with $Cd^{2+}$.
Experimental $T_m$ values are shown by signs: ● for $(T_m)_{21}$, ○ - $(T_m)_{23}$, ★ for $(T_m)_{31}$. -■- calculation values.
▲$_{ss}$, ▲$_{ts}$, ▲$_{ds}$ – structural states as single-, three- and double-stranded molecules, obtained by IR and VCD methods and corresponding to full ion concentrations, A.
---- – Regions corresponding to approximated contents $A_f$ of free ions, used in calculations.

To clear up the reasons of such an essential difference between diagrams with $Ni^{2+}$ and $Cd^{2+}$, it is necessary to examine concentration dependences of binding constants of these ions to all three structures of polynucleotides being used in calculations ($K_A$, $K_U$, $K_{AU}$ and $K_{A2U}$).

The least distinction between binding constants of these ions is observed for the triple chain of polynucleotides ($K_{A2U}$). In this case only Coulomb interactions of ions with negative charges of phosphate groups are possible as well as mutual ion repulsion which decreases binding constants upon the rise of the ion concentration in solution. Therefore, in the case of equal charges, the ion constants $K_{A2U}$ are of equal concentration dependences within the limit of calculation accuracy (Fig. 4).

In contrast to measurements for $Ni^{2+}$, results for $Cd^{2+}$ obtained in our study characterize not the transition temperatures but structural states of polynucleotides, namely 1) fully single-stranded molecules - ▲$_{ss}$, 2) triple-stranded ones + single-stranded poly A - ▲$_{ts}$, 3) – double-stranded ones - ▲$_{ds}$. These states are realized in appropriate concentration and temperature regions.

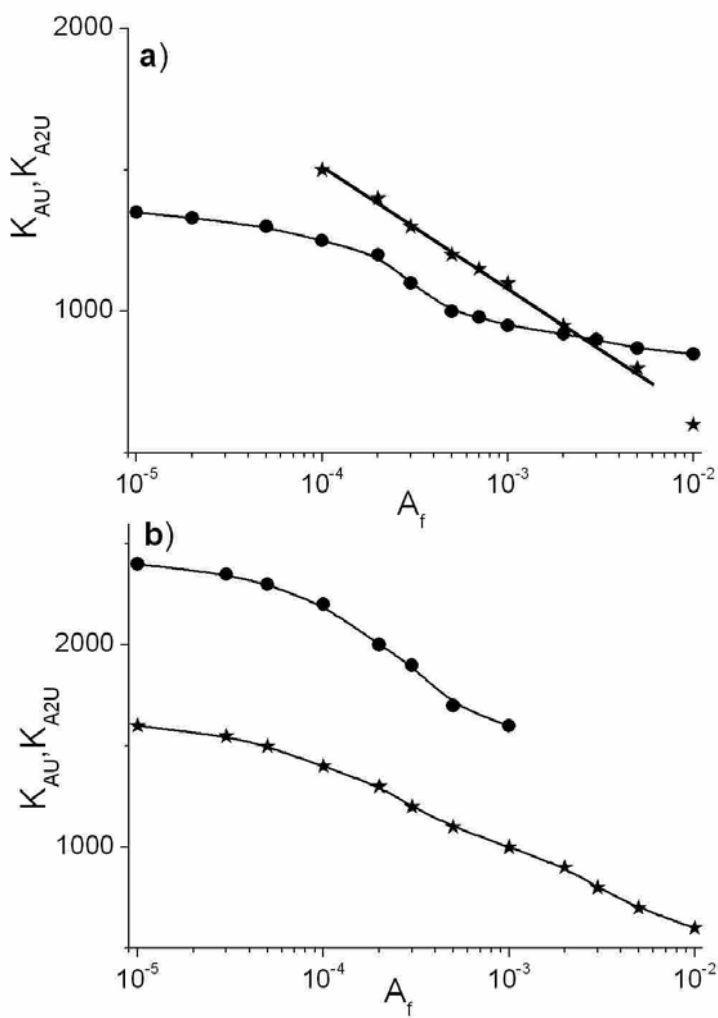

Figure 4. $K_{AU}$ and $K_{A2U}$ dependences on $Ni^{2+}$ (a) and $Cd^{2+}$ (b) concentrations, used in calculations.
● - $K_{AU}$
★ - $K_{A2U}$.

Unlike of the above, constants of the ion binding to the double chain ($K_{AU}$) differ highly essentially (Fig. 4). The difference is conditioned with the possibility of the ion binding to bases, mainly to $N_7$ and, to a somewhat lesser extent, to $N_1$ of adenine, resulting in the macrochelate formation. In comparison with $Ni^{2+}$ ions, $Cd^{2+}$ is of higher affinity to these atoms. As a result, $K_{AU}$ is greater than $K_{A2U}$ (Fig. 4) in spite of the fact that the density of negative charges on the triple chain is higher. This peculiarity of $Cd^{2+}$ has an influence on the form of poly A·poly U diagram and especially on the change of 2→3 transition character, as follows from (6).

The main factor conditioning differences of diagrams for $Ni^{2+}$ and $Cd^{2+}$ is binding of these ions to single-stranded polynucleotides poly A and poly U, especially to the polyadenyl chain ordered at the given temperature.

As noted earlier, in the case of $Cd2^+$ the formation of macrochelate complexes and their subsequent compaction take place at a significantly lower concentration of ions (Fig. 5) than that of $Ni^{2+}$ [7,8]. Besides, $Cd^{2+}$ binding constant $K_A$ is noticeably higher than that for $Ni^{2+}$ and reaches $1.5 \cdot 10^4$ $M^{-1}$ value in the region of molecule compaction, this value being significantly higher than $K_{AU}$ and $K_{A2U}$. Such a high constant magnitude used in calculations agrees well with $K_A$ value [8] obtained earlier at the high degree of the binding site occupation (C~1), the DUVS method being applied in this case.

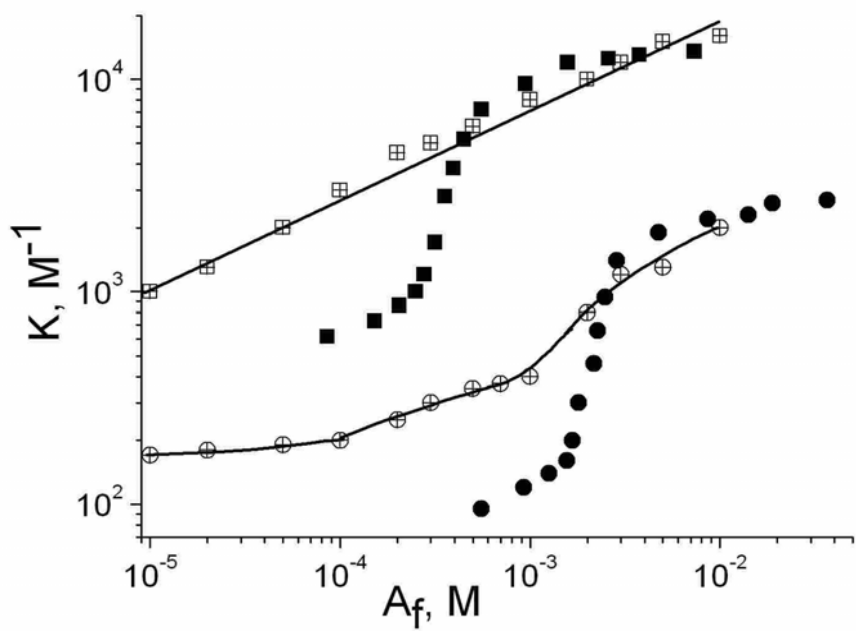

Figure 5. $K_A$ dependences on $A_f$ for solutions with $Ni^{2+}$ and $Cd^{2+}$.
●, ■ – values for $Ni^{2+}$ and $Cd^{2+}$, respectively, obtained by DUVS method and recalculated from linear dependence on occupation C-degree of binding sites of ions and poly A bases [7,8].
⊕, ⊞ - total values of $K_A$ for $Ni^{2+}$ and $Cd^{2+}$, respectively, used in calculations and taking into account ion binding to phosphates groups of polynucleotides.

It follows from (5) and (6) that upon such a significant increase of $K_A$ values $\delta T_{31}$ and $\delta T_{23}$ have to lower sharply (Fig. 3).

Additional contributions into forms of diagrams calculated for 2→1 and 3→1 transitions are made by differences in constants of $Ni^{2+}$ and $Cd^{2+}$ binding to poly U (Fig. 6). As shown earlier [7], $Ni^{2+}$ induces no changes in DUV spectra of poly U, and this is an evidence of its binding only to negative charges on oxygen atoms of phosphate groups. Because of Coulomb ion repulsion upon the binding site occupation, $K_U$ value decreases.

As shown in [10], unlike $Ni^{2+}$ ions, $Cd^{2+}$ binds additionally to poly U bases, forming chelate complexes and ordering poly U molecule.

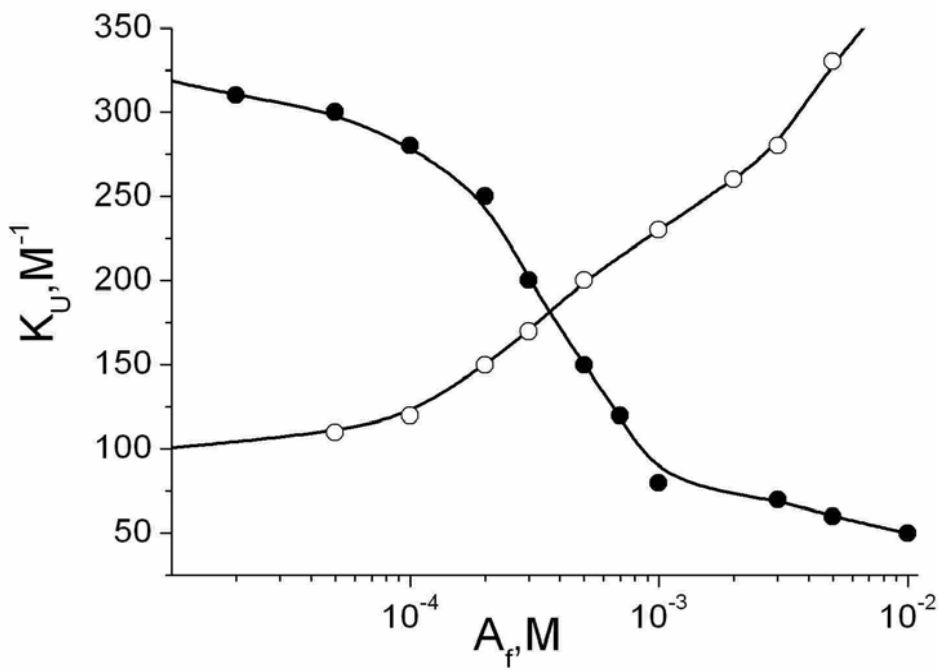

Figure 6. $K_U$ dependences on $Ni^{2+}$ (●) and $Cd^{2+}$ (○) concentrations, obtained upon calculations of diagrams.

$K_U$ values used in calculations (Fig. 6) agree with those obtained in the present work.

## Conclusion

Calculations performed in the present work are of the character of solving the inverse problem aimed at determining parameters in the equilibrium binding theory (the "clip" theory). When comparing calculation results with the experimental data, a solution has been revealed for this inverse problem and dependences of binding constants on cation concentrations have been obtained. Upon calculations, a number of assumptions and approximations have been made. In spite of these operations, main causes have been revealed for differences between polynucleotide phase diagrams in the presence of three ion types representing three classes of metals including such a toxic element as $Cd^{2+}$. In particular, it is shown that one of the main factors conditioning differences in actions of the ions mentioned are values of constants of their binding to the partially ordered molecule of poly A and molecule compaction upon the ion concentration increase. Evidently, to obtain more strict and exact calculation results and conclusions, it is necessary to carry out additional studies on interactions of the ions considered with polynucleotides, using UV-, IR, NMR and other methods. This will permit to take into account all the possible types of ion interactions with polynucleotides. Besides, it is desirable to perform theoretical analysis by new model methods and quantum-chemical calculations. The results may be all-important for gaining a better understanding of the role of ions contaminating the environment. In particular, the strong $Cd^{2+}$ influence on 2→3 transition can explain this metal toxicity.